\documentclass[review]{elsarticle}

\usepackage{lineno,hyperref}
\usepackage[utf8]{inputenc}
\usepackage{graphicx}
\usepackage{xcolor}
\usepackage{setspace}
\journal{Journal of \LaTeX\ Templates}

\biboptions{authoryear}

\begin{document}

\begin{frontmatter}

\title{Zodiacal light observations and its link with cosmic dust: a review}


\author[address1]{Jeremie Lasue}
\cortext[mycorrespondingauthor]{Corresponding author}
\ead{jlasue@irap.omp.eu}
\author[address2]{Anny-Chantal Levasseur-Regourd}
\author[address3]{Jean-Baptiste Renard}

\address[address1]{IRAP, Universit\'e de Toulouse, CNRS, CNES, UPS,  
9 avenue Colonel Roche, FR-31400 Toulouse, France}
\address[address2]{LATMOS, Sorbonne Universit\'e, CNRS, UVSQ,  
Campus Pierre et Marie Curie, 4 place Jussieu, Paris FR-75005, Francee}
\address[address3]{LPC2E, CNRS, Orl\'eans FR-45071, France}


\begin{abstract}

The zodiacal light is a night-glow mostly visible along the plane of the ecliptic. 
It represents the background radiation associated with solar light scattered
by the tenuous flattened interplanetary cloud of dust particles surrounding the Sun and the planets. 
It is an interesting subject of study, as the source of the micrometeoroids falling on Earth, 
as a link to the activity of the small bodies of the Solar System, but also as 
a foreground that veils the low brightness extrasolar astronomical light sources. 

In this review, we summarize the zodiacal light observations that have been 
done from the ground and from space in brightness and polarization at various 
wavelength ranges. Local properties of the interplanetary dust particles 
in some given locations can be retrieved from the inversion of the zodiacal light 
integrated along the light-of-sight.  We show that the current community 
consensus favors that the majority of the interplanetary dust particles detected 
at 1~au originate from the activity of comets. 

Our current understanding of the interplanetary dust particles properties 
is then discussed in the context of the recent results from the Rosetta 
rendezvous space mission with comet 67P/Churyumov-Gerasimenko. 

\end{abstract}

\begin{keyword}
Zodiacal light \sep brightness \sep polarization \sep thermal emission \sep
interplanetary dust \sep cometary dust
\end{keyword}

\end{frontmatter}



\section{Introduction}

\begin{figure}[b]
\begin{center}
 \includegraphics[width=0.8\textwidth]{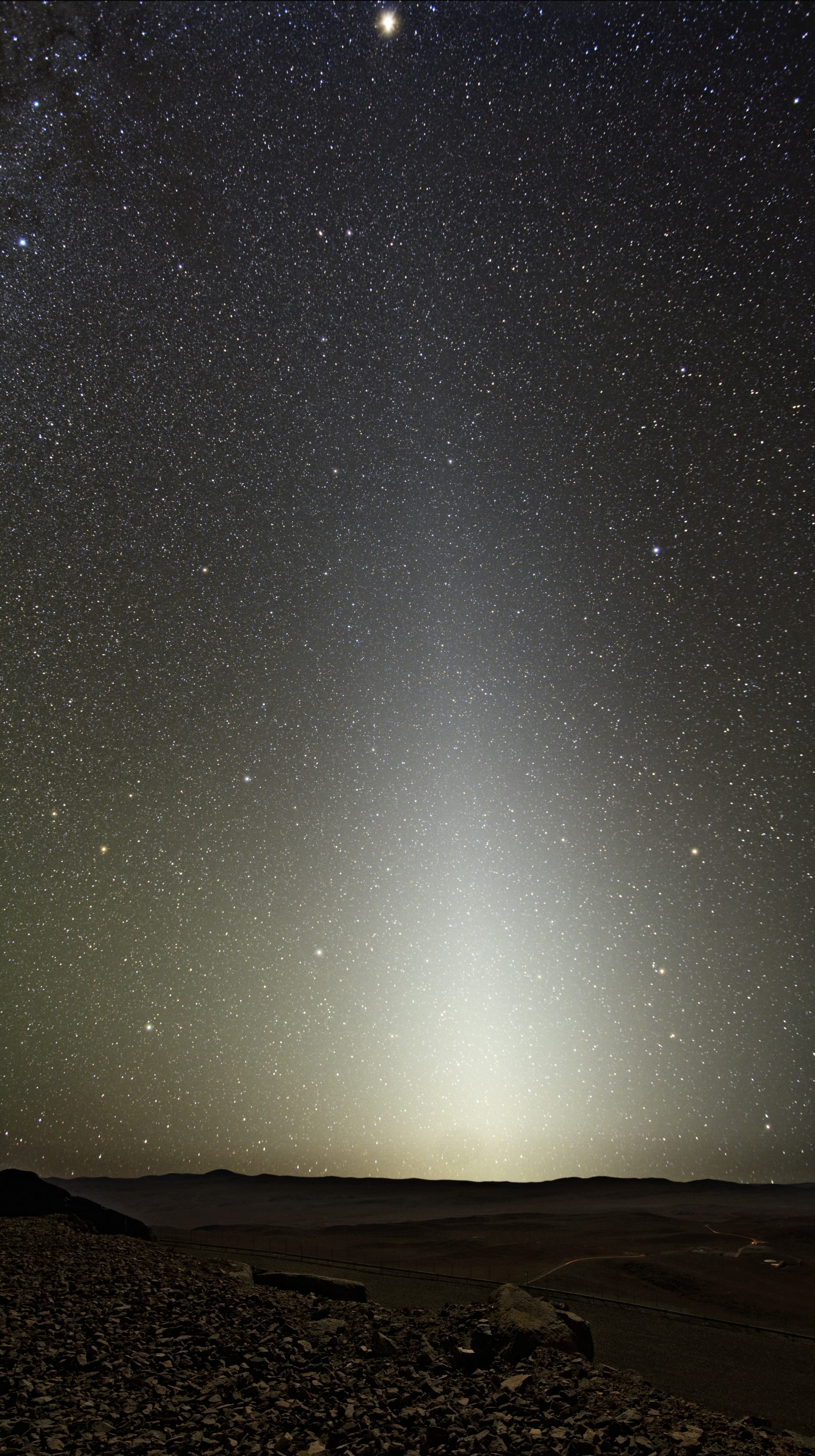} 
 \caption{Zodiacal light seen from Paranal, Chile. The bright planet 
 at the top center is Jupiter (ESO/Y.Beletsky).}
  \label{zod:fig1}
\end{center}
\end{figure}

The zodiacal light is a faint night-glow, brighter as the observer looks along the ecliptic 
and closer to the Sun. 
Its visual appearance usually takes the shape of Sun-colored cones of light visible at dusk 
and at dawn as illustrated in Fig.~\ref{zod:fig1}. 
In fact, the zodiacal light originates from the solar light scattered by the tenuous 
lenticular cloud of interplanetary dust particles surrounding the Sun and 
in which the planets orbit. 
Therefore, the study of its properties is of interest to better understand the origin and 
characteristics of interplanetary dust particles but also to better understand how 
planetary and exoplanetary systems evolve with time. 

\textbf{Figure 1: zodiacal light (ESO/Y.Beletsky)}

The night sky brightness as seen from the ground on Earth 
originates from a combination of i. upper atmosphere airglow, ii. the zodiacal light 
scattered and emitted from the interplanetary dust particles in the solar system, 
iii. the integrated starlight from individual and unresolved stars, iv. the diffuse galactic light 
scattered from interstellar dust particles and emitted from the filamentary cirrus
detected by IRAS \citep{hauser1984} and v. the extragalactic light and background
\citep[see e.g.][]{levasseur-regourd1994}.
The first scientific work on the study of the zodiacal light was published by Cassini 
in the 17\textsuperscript{th} century \citep{cassini1693}. Scientists progressively understood 
that the light was present in  every celestial direction, presented a solar spectrum, 
and was partially polarized. They therefore assumed it to be due to sunlight scattered
by tiny solid particles \citep[e.g.][]{wright1874}.
Quantitative measurements were pursued during the 20\textsuperscript{th} century with a view
to better understand its constituting particles 
and to characterize this foreground signal to improve 
extrasolar astronomical observations.
Extensive historical notes on the study of the zodiacal light can be found in 
e.g. \citet[][]{fechtig2001, levasseur-regourd2001b}.

The cloud's brightness comes  principally from scattering in the visible and 
by thermal emission in the infrared and so can be studied in both 
wavelength ranges. 
The most complete reviews of its properties were 
published almost 20 years ago in the report from the  
\textit{IAU Commission 21: Light of the night sky}
``The 1997 reference of diffuse night sky brightness'' \citep{leinert1998}
and in the ``Interplanetary dust'' book published in 2001 \citep{grun2001a}.
Since then, a number of programs and space missions have improved 
the available measurements of the zodiacal light 
and our knowledge of the properties of the interplanetary 
dust particles that constitute it, 
like the space missions Solar Mass Ejection Imager (SMEI)
in the visible \citep{buffington2016}
and AKARI in the infrared \citep{pyo2010}. 
In this work we will review the properties of the zodiacal light from observations 
and describe the properties of the interplanetary dust particles 
that can be deduced from them. 

\section{Integrated observations of the zodiacal light along the line of sight}

\subsection{Intensity observations}

The initial studies of the zodiacal light originated from a need to better understand
the solid matter environment surrounding the Earth, together with the fact 
that this glow was a source of noise to the study of faint objects and extended 
sources \citep[see e.g.][]{dufay1925, dumont1965, leinert1998}. 
The night sky brightness includes atmospheric glow, interplanetary dust light 
scattering and thermal emission, as well as galactic and extra-galactic point or 
diffuse sources. 
Long term surveys can disentangle these different signal sources 
to retrieve each component separately. The study of the zodiacal light 
necessitates year-long surveys to obtain a full assessment of its geometry
and determine possible time variations of the dust cloud. 

\begin{figure}[b]
\begin{center}
 \includegraphics[width=0.8\textwidth]{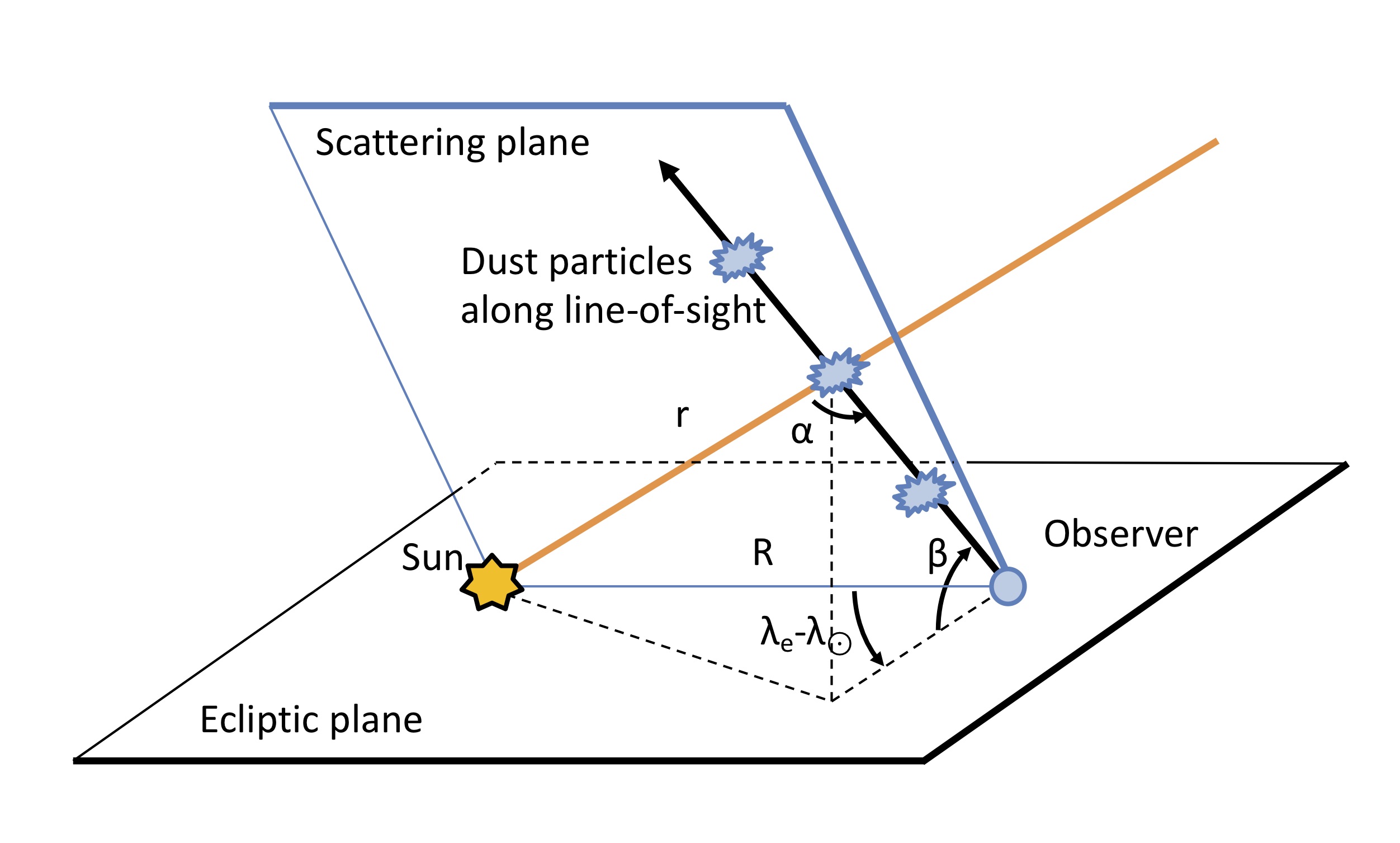} 
 \caption{Zodiacal light geometry of observation in helio-ecliptic coordinates.}
  \label{zod:fig2}
\end{center}
\end{figure}

The zodiacal light intensity, $I$, depends on the position of the observer
along the terrestrial orbit, the wavelength and 
the viewing direction with respect to the Sun. The dust distribution around 
the Sun essentially follows a mirror symmetry with respect to a surface close 
to the ecliptic. 
This symmetry surface is close to a plane, 
but may diverge somewhat from it 
and is inclined by $i = 3.0^{\circ} \pm 0.3$ from the ecliptic plane as
measured by the Helios probes 
between Earth and Venus orbits \citep{leinert1980}. 
Helio-ecliptic coordinates, illustrated in Fig.~\ref{zod:fig2},
are generally suitable to describe the zodiacal light with the ecliptic 
latitude $\beta$ and the helio-ecliptic longitude 
($\lambda_e-\lambda_{\odot}$), where 
$\lambda_e$ is the ecliptic longitude. The phase angle  of observation
is noted  $\alpha$.
As shown in Fig.~\ref{zod:fig2}, one needs to keep in mind that any signal observed 
in a given direction corresponds to an integrated signal along the line of sight. 

\textbf{Figure 2: geometry of observation}

\begin{figure}[b]
\begin{center}
 \includegraphics[width=0.8\textwidth]{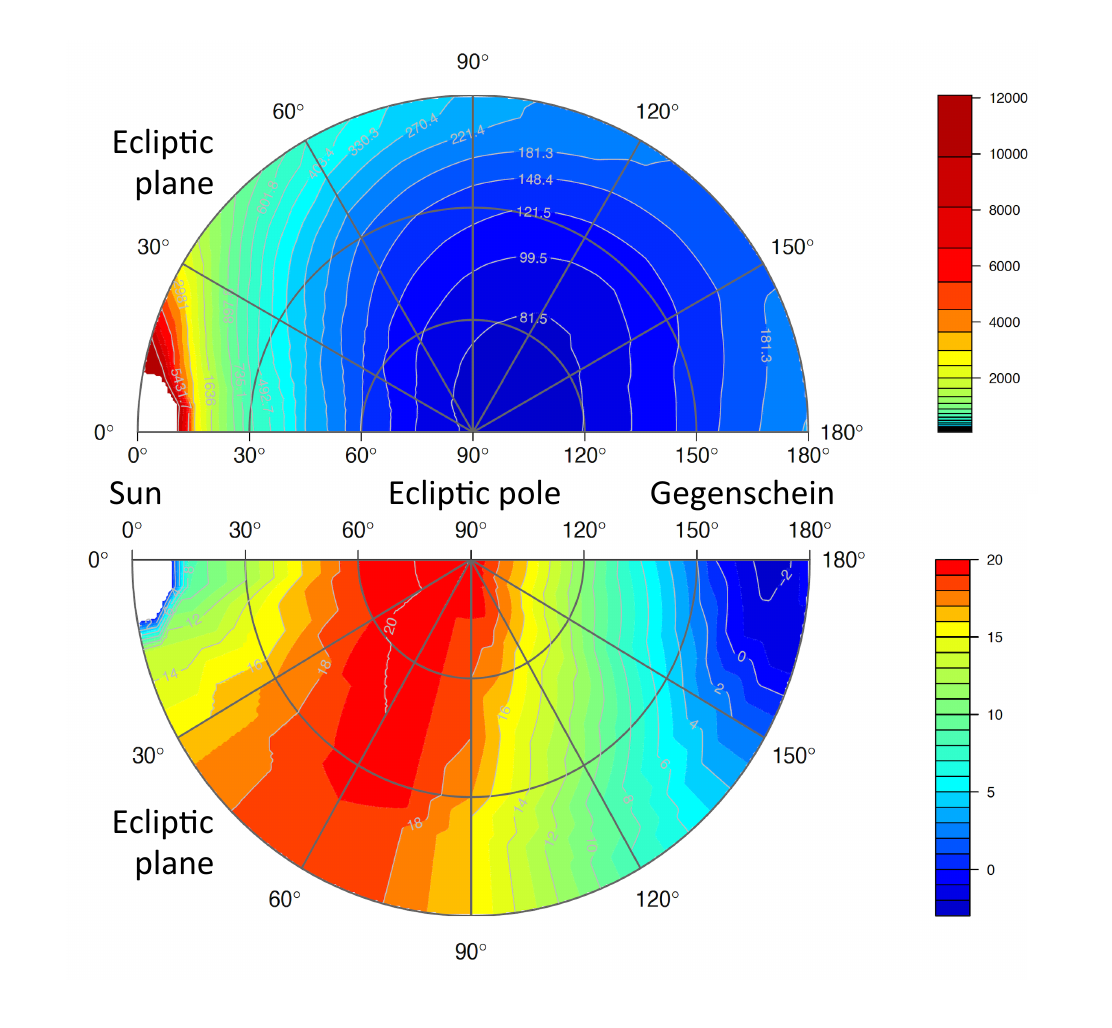} 
 \caption{Upper part: Map of the zodiacal light brightness 
 ($10^{-8}$~W~m$^{-2}$~sr$^{-1}~\mu$m$^{-1}$) at 550 nm.
 Lower part: Map of the zodiacal linear polarization (in percent).
The outer circle represents the ecliptic, the point at $0^{\circ}$ is the Sun, 
the central point is the ecliptic pole, and the near $180^{\circ}$ region is 
the Gegenschein, with an increase in brightness by backscattering, 
and a low and even negative linear polarization between $160^{\circ}$ and $180^{\circ}$. 
Values are corrected for oscillations induced by the slight inclination of 
the zodiacal light symmetry plane and the Earth's orbit eccentricity. 
Based on data from \citet{levasseur-regourd1980, levasseur-regourd1996} and 
\citet{leinert1998}; updated from \citet{lasue2015a, levasseur-regourd2019a}.}
  \label{zod:fig3}
\end{center}
\end{figure}

In the early years of space exploration a better understanding of the dust 
material surrounding the Earth was sought, and systematic observations of 
the zodiacal light were performed. 
High altitude surveys were performed with narrow filters to limit airglow 
contamination over several years in Hawaii  \citep{weinberg1964a}. 
At the Canary Islands  \citep{dumont1965} devised a method to remove the night airglow 
using a correlation between the 557.7~nm O I line and the atmospheric continuum 
observed on multiple points over the celestial sphere.
Other astronomers' surveys were also compiled in  \citet{leinert1975} and \citet{fechtig1981}, 
and include data obtained by \citet{weinberg1964a, blackwell1967a, vandenoord1970, 
sparrow1972, leinert1976}.
Fig.~\ref{zod:fig3} represents the zodiacal light signal averaged over a year
in intensity, $I$, and linear degree of polarization, $P$ (which will be discussed in Section~2.2),
on a diagram representing a quarter of the celestial sphere. The values presented
correspond to the unification of the survey measurements published 
in tabular form in the review of \citet{leinert1998}. 
The outer circle represents the ecliptic. 
The minimum brightness is obtained at a point close to the ecliptic 
pole coordinates \citep{dumont1975a, dumont1975}. 
Taking into account the fact that the zodiacal light has a solar spectrum, its intensity $I$ 
at 550~nm is then found to be close to $2.4 \times 10^{-5}$~W~m$^{-2}$~sr$^{-1}~\mu$m$^{-1}$ 
at $30^{\circ}$ solar elongation in the ecliptic plane (which outshines the brightest part of 
the Milky Way). It reaches a minimum of $7.6 \times 10^{-7}$~W~m$^{-2}$~sr$^{-1}~\mu$m$^{-1}$
in the vicinity of the ecliptic pole. In the ecliptic plane, a minimum of 
$1.8 \times 10^{-6}$~W~m$^{-2}$~sr$^{-1}~\mu$m$^{-1}$ is reached at around 
$140^{\circ}$, which is still 2.3 times higher than the ecliptic pole value. 
The resolution of the data is between $5^{\circ}$ and $15^{\circ}$. 
The relative uncertainty in the intensity data remains below 5\%.
One notices the continuous decrease of $I$ as the viewing direction 
moves away from the Sun, except for the direction directly opposite 
the Sun, named the \textit{gegenschein}, where an increase in brightness 
is due to light backscattering and $I$ reaches a value about 1.3 times larger than 
the ecliptic minimum. 

\textbf{Figure 3: Brightness and polarization maps}

Meanwhile, several space mission observations were also obtained with 
the added advantage of not being contaminated by the atmospheric night-glow, 
but limited in viewing directions and mission time span.
The characteristics of the dust were assessed in different zones of the Solar System 
thanks to their mobility. 
The D2A satellite around the Earth measured the zodiacal light in a plane 
perpendicular to the Sun-Earth direction. Its results 
confirmed the validity of ground-based measurements, and gave some evidence
for the presence of interplanetary dust heterogeneities \citep{levasseur1973}. 
The Helios 1 and 2 probes showed an increase of $I$ with decreasing 
heliocentric distance within the ecliptic plane \citep{leinert1981, leinert1982}. 
This lead to the assumption that the dust particles population density, $n$, was 
increasing towards the Sun as $n(r)\propto r^{-1.3}$, where $r$ is the radial
distance to the Sun. 
Toward the outer planets, observations were performed by  Pioneer 
10 and 11 \citep{hanner1976}, establishing that the zodiacal light intensity 
decreases away from the Sun and becomes non detectable after
3~au \citep{matsumoto2018a}.
These measurements confirmed and extended the surveys that were obtained
from the ground. 

For an observer located on Earth's orbit, the zodiacal light intensity will vary 
depending on the distance to the Sun due to the Earth's orbit eccentricity. 
Once this effect is corrected for, $I$ observed at 1~au remains consistently 
stable within 1.5\% over a whole solar cycle \citep{levasseur-regourd1996}. 
The remaining annual residual oscillations originate in a slight
inclination of the zodiacal cloud's symmetry plane with respect to the ecliptic.
At 1~au, this symmetry plane presents an ascending node at $95^{\circ} \pm 20^{\circ}$ 
with an inclination of $1.5^{\circ} \pm 0.4^{\circ}$ \citep{dumont1978}. 
The symmetry surface of the zodiacal cloud is possibly warped far away from 
the Earth, as an inclination of  about $3^{\circ}$ was inferred from Helios measurements
\citep{leinert1980} and a dust ring associated with the orbit of Venus detected \citep{leinert2007}.

\begin{figure}[t]
\begin{center}
 \includegraphics[width=0.8\textwidth]{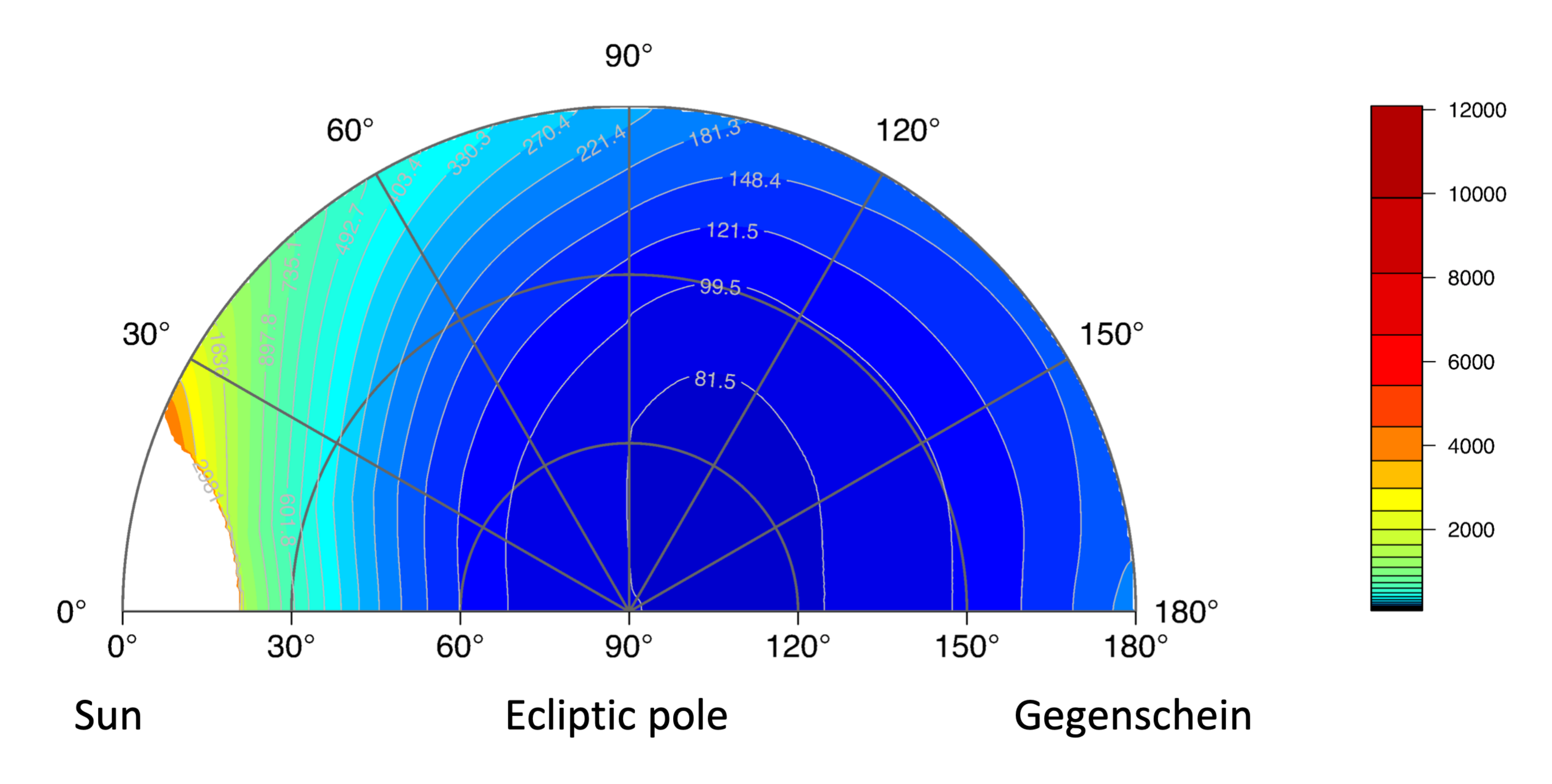} 
 \caption{Most recent map of the zodiacal light brightness 
 ($10^{-8}$ W m$^{-2}$ sr$^{-1}$ $\mu$m$^{-1}$). 
Brightness levels are comparable to those presented in 
the upper part of Fig.\,\ref{zod:fig3}, while the resolution is significantly improved.
The outer circle represents the ecliptic, the point at $0^{\circ}$ is 
the Sun, the central point is the ecliptic pole, and the 
near $180^{\circ}$ region is the Gegenschein. 
Based on data from \citet{buffington2016} and adapted from \citep{levasseur-regourd2019a}.}
\label{zod:fig4}
\end{center}
\end{figure}

A re-analysis of Weinberg's observations from Hawaii has recently yielded a new intensity map 
with a $2^{\circ}$ resolution from which the ascending node position gives around $80^{\circ}$ 
while the inclination of the cloud's symmetry plane is around $2^{\circ}$ \citep{kwon2004a}. 
Later on, the whole sky daily photometric observations by  the Solar Mass Ejection Imager (SMEI) 
space mission were used to generate an even more precise and resolved zodiacal light 
map \citep{buffington2016}. 
After image processing to remove the brightest stars, 
daily zodiacal maps with a resolution of  $0.5^{\circ}$ were calculated over the 8.5 years 
of the satellite mission  (from 2003 to 2011). 
An average zodiacal map is shown in Fig.~\ref{zod:fig4} in
the same reference frame as Fig.~\ref{zod:fig3}. 
Based on these measurements, an upper limit to the zodiacal light intensity changes 
of 0.3\% was obtained for the time period of observations of 8.5 years. 
Using the same satellite data, \citep{buffington2009}
determined that the {\it gegenschein} is always located at the anti-solar point, but varies by 
$\approx 10\%$ of its intensity over time, with a portion of the variation repeating seasonally. 
Similarly, the latest high-resolution (5' per pixel) survey of the {\it gegenschein} using a 
wide-angle camera showed that the maximum scattered intensity is consistently located 
at the anti-solar point. \citet{ishiguro2013}  fitted a model to the {\it gegenschein} geometry
and constrained a very low albedo of 0.06 for the particles located in that direction. 

\textbf{Figure 4: Buffington zodiacal map}

Local short-term enhancements in the zodiacal light brightness are also detected
and were assumed to be due to possible meteoroid streams
\citep[e.g.][]{levasseur1973}. Since then the discovery of faint cometary trails in the 
infrared by the Infrared Astronomical Satellite (IRAS) \citep{sykes1992} and 
the Spitzer telescope \citep{reach2007}
as well as linear dust features in the optical domain 
\citep{ishiguro1999, yang2012} have clearly demonstrated
the existence of many local enhancements in dust density due to the activity and impacts 
amongst the Solar System small bodies population 
\citep[See e.g. Fig. 4 in][]{levasseur-regourd2019a}.

Extensive descriptions of the results and methods of zodiacal light surveys can be found 
in the reviews previously published 
\citep[see ][]{leinert1975, weinberg1978, leinert1990, leinert1998, 
levasseur-regourd2001b, mann2004, lasue2015a}

\subsection{Polarimetric properties}

\subsubsection{Degree of linear polarization observations}

As light scattered by an optically thin cloud of dust particles, the zodiacal light 
presents a systematic linear polarization, $P$, defined as the ratio of the 
difference between the light intensity perpendicular, $I_{\perp}$, and parallel, 
$I_{\parallel}$, to the scattering plane
and the total intensity $P=\frac{I_{\perp}-I_{\parallel}}{I_{\perp}+I_{\parallel}}$.
The linear polarization $P$ can reach values 
as high as 20\% in the integrated line of sight. $P$ is normalized and positive if 
the electric field polarization direction is perpendicular to the scattering plane as shown 
in Figure~\ref{zod:fig2}. Observations demonstrated that the majority of the night 
sky presents a positive linear polarization signal \citep{weinberg1985}.

Fig.~\ref{zod:fig3} bottom, represents the values of polarization annually averaged 
over the sky in differential ecliptic coordinates based on data from \citet{dumont1976}
and Table 18 published in \citep{leinert1998}. 
The error in the measured polarization is estimated to be $\approx 2\%$. 
The linear polarization is clearly dependent to the first order on the helio-ecliptic 
longitude. It decreases to zero towards the Sun, in the forward scattering direction, 
and reaches a maximum value of $\approx 20\%$ near $60^{\circ}$ of helio-ecliptic
longitude and near the ecliptic pole.  
The linear polarization signal then presents negative values around the 
\textit{gegenschein} with a minimum of about -2\% near $165^{\circ}$ of elongation
similar to the backscattering signal obtained with light scattered 
by irregular dust particles of sizes of the same order of magnitude as the light's 
wavelength, $\lambda$  \citep{levasseur-regourd1997, lumme1997}.
While the negative value of polarization in that region of the sky may be difficult 
to correctly assess as discussed in \citep{dumont1975}, the presence of  a negative polarization 
region surrounding the \textit{gegenschein} was confirmed from measurements 
done on the ground \citet{weinberg1968}, from sounding rockets \citet{wolstencroft1967} 
as well as from balloons \citet{frey1974}, making it a robust detection.
The inversion angle corresponds to $15^{\circ} \pm 5^{\circ}$, 
with a slope at inversion of ($0.2 \pm 0.1$) percent per degree.
Therefore the integrated linear polarization variation of the zodiacal light 
in the ecliptic plane follows variations similar to the ones of typical 
polarimetric phase curves of the Solar System small bodies surfaces 
(e.g., regolith on the Moon and on asteroids) or dust clouds (e.g., dust in cometary comae). 
This trend is typical of scattering by irregular particles, with a size greater than  the wavelength
(\citep{belskaya2019}).

\begin{figure}[t]
\begin{center}
 \includegraphics[width=0.8\textwidth]{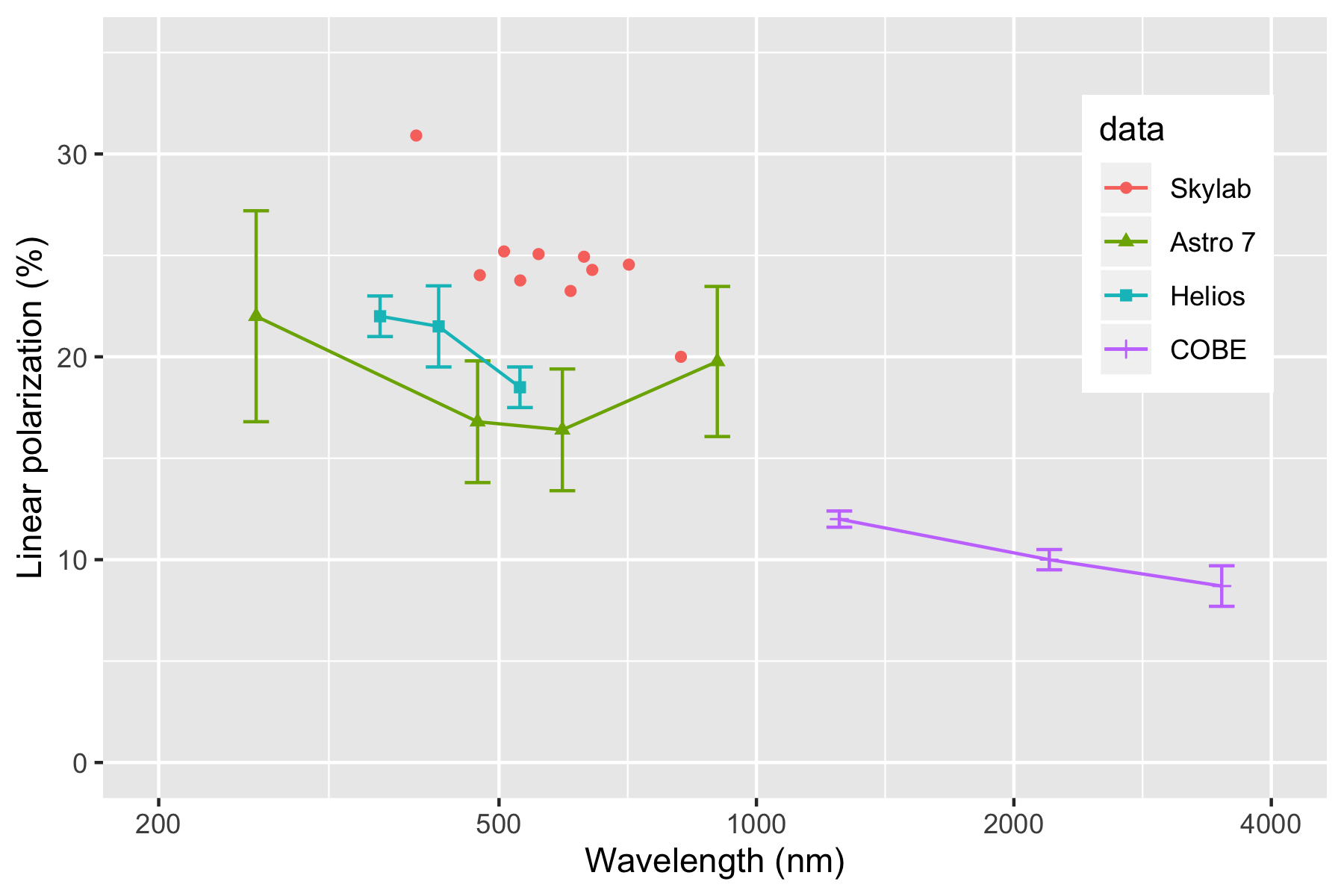} 
 \caption{Wavelength variation of the linear polarization, $P$, in different directions of the sky. 
 Skylab at the North Celestial Pole \citep{weinberg1980}; 
 rocket Astro 7 ($\lambda_e - \lambda_{\odot}=25^{\circ}$) \citep{pitz1979}; 
 Helios ($\lambda_e -  \lambda_{\odot}=90^{\circ}$) \citep{leinert1982}; 
 COBE ($\lambda_e -  \lambda_{\odot}=90^{\circ}$) \citep{berriman1994}.}
  \label{zod:fig5}
\end{center}
\end{figure}

Polarimetric observations have been done at different wavelengths and 
towards different directions of the sky as summarized in Fig.~\ref{zod:fig5}. 
There are no strong trends in the visible range, 
from 450~nm to 800~nm, but near infrared observations from 1~$\mu$m to  4~$\mu$m
by COBE could indicate a decreasing trend towards higher wavelengths \citep{leinert1998}. 
This could be due partly to contamination by the thermal dust emission, 
which starts at about 3~$\mu$m \citep{berriman1994}. 
Also, because the observations summarized in Fig.~\ref{zod:fig5}
probe different regions of the zodiacal cloud complex, the scatterers must have different origins. 
As different spectral polarimetric trends are seen 
for different types of asteroidal surfaces \citep{belskaya2017} and cometary dust clouds
\citep[see e.g.][]{levasseur-regourd1996a, hadamcik2003}
further more precise observations may indicate potential genetic links to the dust sources.

\textbf{Figure 5: Spectral gradient for polarization}

\subsubsection{Solar F-corona observations}

Closer to the Sun, the zodiacal cloud starts to merge with the solar corona. 
The brightness of the solar corona originates  mainly from three physical processes:
i. Thomson scattering by free electrons (K-corona, major component at $<4$ solar radii), 
ii. line emission from ions (L-corona), 
and iii. light scattering from dust particles (F-corona). 
The F-corona emission dominates at the largest distances and can be separated from 
the other emissions by spectrometric or polarimetric analyses. 

The size range of observed particles may change in the F-corona compared to
the zodiacal light,  causing a change of their average optical properties.
Typical sizes of the particles are still expected to remain larger than $1~\mu$m 
\citep{mann2004}.
The change in single particle scattering properties at small angles, however,
is expected to change the radial slope of the F-coronal brightness compared to
the zodiacal light. Combining overall measurements, it is expected that 
the brightness varies as $R^{-2.5}$ at the solar equator and $R^{-2.8}$ at
the poles, with $R$ the radial distance to the Sun 
\citep[see][ and references therein]{levasseur-regourd2001b}.

Color trends are not evidenced from measurements at wavelengths
below $1~\mu$m \citep{kimura1998}. At larger wavelengths, 
the F-corona brightness presents a reddening. The
color is expected to vary within the corona as a result of the superposition of
light scattering and thermal emission effects \citep{mann1993}.
Possible dust rings around the Sun in the region where material sublimation 
may occur were anticipated due to the balance
between the inward Poynting-Robertson drag and the outward radiation pressure. 
Observations of  the 1991 solar eclipse have shown no evidence for 
the existence of a dust ring around the Sun 
\citep[see][and references therein]{levasseur-regourd2001b}.

Typical classical coronal models expect the F-corona polarization to decrease and tend 
to zero towards the Sun \citep{blackwell1967}.
Polarimetric observations of the Solar corona \citep{skomorovsky2012, burkepile2017} and 
data obtained by Clementine during solar occulations by the Moon \citep{hahn2002} have 
confirmed the variations measured in previous works 
\citep[see reviews by ][]{levasseur-regourd2001b, mann2004}.

\subsubsection{Circular polarization}

The light scattered by the interplanetary dust cloud is, to the first order, 
linearly polarized. However, multiple scattering, 
partial alignment of the dust particles, or even optically active materials can introduce 
a circular polarization component in the scattered light \citep[see \textit{e.g.}][]{mishchenko2002}. 
Measurements of the circular polarization have been attempted from space 
and on the ground but remain inconclusive due to the very low level of signal expected
\citep{wolstencroft1967, staude1972, wolstencroft1972}. 
Observations do not indicate circular polarization values in excess of 0.1\% for most of the sky
except for a possible signal in excess of  0.5\% around a helio-ecliptic longitude 
of $240^{\circ}$ \citep{wolstencroft1972}.
In the case of the zodiacal cloud, one expects the major processes to explain a possible circular 
polarization to be either: i.particle alignment \citep{lazarian2015, lasue2015a}, 
or ii. optically active materials. 
Both processes have been suggested to explain the very low circular polarization measured
for comets with recent measurements by Rosetta favouring a particle alignment 
\citep{rosenbush2007, kolokolova2016}. 
Further attempts to observe the zodiacal light circular polarization may be of importance 
as  it could constrain the origin of the particles and specific dynamic 
properties in the solar magnetic field. 

\subsection{Spectral properties}

In the visible part of the spectrum, the zodiacal light shows a solar spectrum. 

Looking at the spectral properties of the zodiacal light integrated along the line-of-sight
at larger wavelengths, 
a number of interesting measurements have been performed, 
especially with space missions such as IRAS, COBE, ISO, AKARI, 
WISE and Planck,
to deduce the physical properties and the dynamics 
of the dust particles in the interplanetary dust cloud. 

First, many infrared measurements have put into evidence the presence 
of dust trails and rings associated with the impacts of asteroids and 
the activity of comets \citep[e.g.][]{dermott1984, sykes1988, reach1997}
The spectral analysis of the Doppler shifted scattered solar light Fraunhofer lines
gives insight into the dynamics of the dust particles constituting 
the zodiacal cloud.  \citet{ipatov2008} have shown the measurements to be consistent 
with a majority of the dust particles associated with cometary activity close to 
the Sun ($<1$~au). 

Then, a red spectral gradient of the scattered light is determined by many measurements
in the range from 200~nm to 2~$\mu$m, 
as shown by COBE \citep{berriman1994}, and
as summarized in \citep[][and references therein]{leinert1998}. 
The spectral gradient can be calculated to be $S=8.5 \pm 1.0 \% / 100$~nm at 460~nm
\citep{yang2015}. Combining this value with the albedo of $A=0.06 \pm 0.01$ 
measured at the same wavelength  in the direction of the \textit{gegenschein} by 
\citet{ishiguro2013} shows that such particles scattering properties 
are most consistent with the properties of primitive small bodies such as comets 
or D-type asteroids \citep{yang2015}.

Measurements (from balloons, rockets and satellites) of the thermal 
emission originating from the zodiacal cloud have also been performed. 
The zodiacal thermal emission, which gradually takes over the solar-spectrum by about 
2~$\mu$m, is actually the most prominent source of brightness in the 5 to 100 micrometer 
range, at least away from the galactic plane \citep[e.g.][]{leinert1998}. 
For even larger wavelengths, the sky brightness gets dominated by 
the interstellar medium, the cosmic diffuse infrared background and 
finally the cosmic microwave background. 

This thermal range can be modelled using blackbody emission curves for 
specific mixtures of materials (like silicates, amorphous carbon and graphite). 
A change of slope in the spectrum at about $150~\mu m$ is interpreted 
to represent a change in the dust size distribution at a radius of about $30~\mu m$ 
\citep{fixsen2002},
which is consistent with the size distribution variations measured in situ \citep{grun1985}.
Comparison of the model with the data by \citet{fixsen2002} indicates that 
replenishment of the interplanetary dust is $\approx 10^{11}~kg~yr^{-1}$ from asteroids and comets.
Dust emission rate from asteroids impacts and activity appears to be lower and thus, 
a significant contribution from comets  would be needed to replenish 
the interplanetary dust cloud at 1~au.

The work by \citet{reach2003} with the ISO satellite data also indicates 
a dust distribution dominated by  large ($>10~\mu m$ radius), 
low-albedo ($<0.08$), rapidly-rotating, gray particles at 1~au from the Sun.
Spectral signatures of amorphous and crystalline silicates are also detected 
in the infrared from 9 to $11~\mu m$ indicating also the presence of 
small particles similar to the ones ejected by active comets, such as  
C/1995 O1 Hale-Bopp comet, and to proto-planetary disks emission
spectra. 
Similar spectral signatures of crystalline silicate features were obtained 
with the AKARI spacecraft \citep{ootsubo2009}.

At even larger wavelengths, in the microwave domain, Planck measurements
have shown that the zodiacal cloud is a minor correction to the Cosmic Microwave 
Background emission \citep{ade2014}.

\section{Interplanetary dust local properties from local inversion}

So far, we have discussed only the properties of the zodiacal light, 
which is an integrated signal (Fig.~\ref{zod:fig1}) 
along the line of sight of the observer (Fig.~\ref{zod:fig2}). 
However, techniques have been tentatively developed to invert this signal in some cases,
thus giving information on the local properties of the brightness
and the polarization of the zodiacal light. 
This is what we describe in this section. 

\subsection{Inversion method}

The dust in the zodiacal cloud cannot be assumed to have the same 
optical properties everywhere and in all directions.
In the data integrated over the line of sight, contributions from light scattering 
by dust particles corresponding to different distances to the Sun, 
different observing phase angles and 
with likely different physical properties are mixed. 
In order to obtain specific properties of the dust particles, it is mandatory 
to invert the integrated intensity and polarization. 
Knowing that the zodiacal cloud is seen to be relatively homogeneous 
and in an apparent steady state, which is consistent with the mixing of injected
dust particles and their short lifetimes 
\citep{campbell-brown2008,nesvorny2011,yang2018}, it is possible to invert
the light integrated over the line-of-sight to deduce local scattering properties 
of the dust particles.
Local rigorous inversion and local inversions through mathematical methods are, 
together with their main results, summarized in \citet{levasseur-regourd2001b}, 
section 4.

Assuming that the zodiacal cloud has a plane of symmetry, within which 
the observer is located, and that the density of dust particles varies as a power-law 
with the heliocentric distance and is in an equilibrium state, 
a rigorous inversion of the zodiacal light intensity 
integrated along the line-of-sight can be derived from Earth-based observations 
as a function of the phase angle at 1~au \citep{dumont1973}.  
Given the local values for the light intensity polarized parallel and perpendicular
to the scattering plane, the associated local linear polarization can be directly 
calculated.
This method is used to determine that at a phase angle $\alpha=90^{\circ}$ 
at the symmetry plane, one cubic kilometer of interplanetary space scatters sunlight with 
an intensity that is $4 \times 10^{-34}$ times smaller than the solar brightness \citep{schuerman1979}. 
The local linear polarization is calculated to be about 
$0.30 \pm 0.03$ \citep{levasseur-regourd1990}.

\begin{figure}[t]
\begin{center}
 \includegraphics[width=0.8\textwidth]{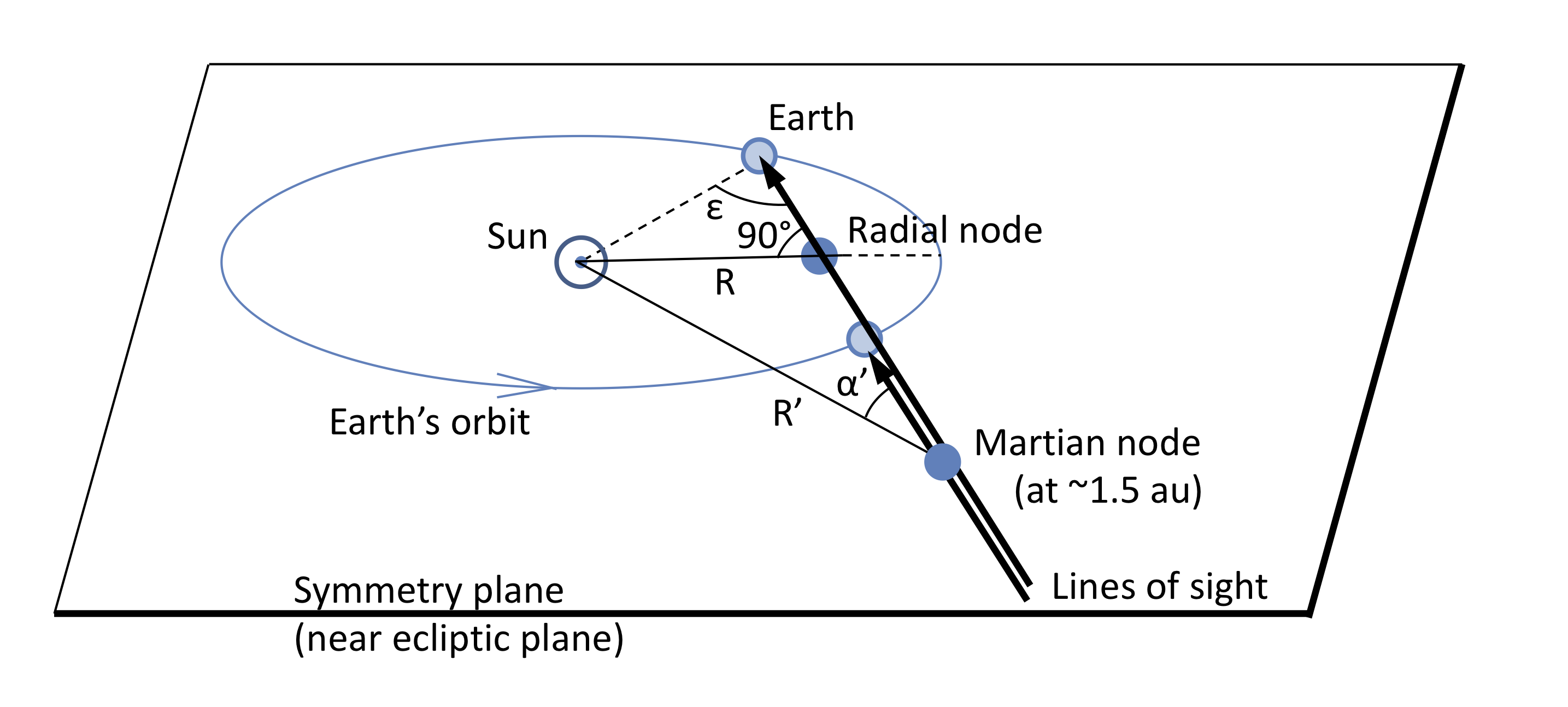} 
 \caption{ Geometry of observations for inversion of the zodiacal light, 
 for lines of sight in the symmetry plane (near the ecliptic) of the zodiacal cloud 
 \citep[Figure adapted from ][]{renard1995}.
 Note that these regions correspond to places where mathematical inversion 
 of the line-of-sight integral is possible with minimal assumptions, and not where 
 measurements were made. See text for additional details.}
  \label{zod:fig6}
\end{center}
\end{figure}

\citet{dumont1983, dumont1985} developed 
the method of the ``nodes of lesser uncertainty'' in order 
to invert the measurements outside the Earth's orbital path. 
This method assumes rotational symmetry for the symmetry plane of the zodiacal cloud,
a radial power-law decrease in the dust density from the Sun  
with a quasi-equilibrium state of the cloud density on the time-scale of a few weeks. 
Knowing that the local intensity scattering function is relatively smooth 
and regular, it can be represented by continuous mathematical
functions (e.g. polynomials, or Fourier series) with five parameters 
out of which one is free. 
By using two lines of sight with solar elongations $\epsilon$ and $180^{\circ}-\epsilon$, 
these mathematical functions present two regions where the local brightness varies over a very 
limited range allowing to retrieve its value with a small uncertainty in those positions. 
These mathematical regions correspond to : i. the so-called``radial node'', located near the helio-ecliptic meridian
with a variable solar distance, $R$, below 1~au and a constant phase angle, $\alpha=90^{\circ}$, 
and ii. the so-called ``martian node'' located at a quasi-constant solar distance 
of $R \approx 1.5~au$ but covering a phase angle range from $0^{\circ}$ to $42^{\circ}$ 
giving information on the backscattering function of the dust particles. 
Figure~\ref{zod:fig6} illustrates the method and the position of the nodes. 
Note that the regions were named for illustration purposes but do not correspond to
regions where measurements were done. They were retrieved with limited assumptions
by mathematical inversion of the line of sight integral. 
Some further methods of inversion have also been devised to retrieve local contributions in the 
plane tangential to the Earth's orbit, out of the ecliptic \citep{renard1995} and 
at other positions in the solar system \citep{schuerman1979}. Improvements in the zodiacal 
light data collection will lead to further work on the local intensity and polarization 
characterization with a better spatial and angular resolution.

\textbf{ Figure 6: Observation geometry for zodiacal light inversion. }

A summary of the local properties of the interplanetary dust (intensity, polarization, temperature, 
albedo, number density) in the symmetry plane as inverted from the data and at 
$\alpha = 90^{\circ}$ is presented in Table~\ref{zod:tab1}. The local values for the 
 albedo at a given phase angle (around 90 degrees)
is retrieved by combining the local values of the flux emission, 
the temperature and the intensity \citep{hanner1981}.

\begin{table}
\centering
\begin{tabular}{l | c | r | r }
 & Value at 1 au &  Power law exponent & Domain in au \\
 &  &  (also called gradient) &  \\\hline
Intensity & $\approx 23 \times 10^{-7}$  & $-1.25 \pm 0.02 $ & 0.5 to 1.4 \\
 &    W~m$^{-2}$~sr$^{-1}$~$\mu$m$^{-1}$~rad$^{-1}$ &  &  \\
Degree of &  $30 \pm 3$ \% & $+0.5 \pm 0.1$ & 0.5 to 1.4 \\
linear polarization & &  &  \\
Temperature & 250~K$\pm 10$~K & $-0.36 \pm 0.03$  & 1.1 to 1.4 \\
Albedo & $0.07 \pm 0.03$ & $-0.34 \pm 0.05$ & 1.1 to 1.4 \\
Space density & $10^{-19}$~kg~m$^{-3}$ & $ -0.93 \pm 0.07$ & 1.1 to 1.4
\end{tabular}
\caption{\label{zod:tab1} Variation of the local properties of the dust in the symmetry plane. 
The properties are described as a function of solar distance with a power law assumption. 
The optical properties (linear polarization, albedo, and intensity in 
W~m$^{-2}$~sr$^{-1}$~$\mu$m$^{-1}$~rad$^{-1}$ at 550~nm) 
are retrieved at $\alpha = 90^{\circ}$ \citep{levasseur-regourd1996, levasseur-regourd2001b}.
The density of the particles is inferred from the inverted values in intensity and polarization.}
\end{table}

\subsection{Heliocentric variations}

At the radial node, the inverted values for the local variation in intensity and polarization 
are presented in Fig.~\ref{zod:fig6} at $\alpha=90^{\circ}$ as a function of the heliocentric 
distance. 
The figure is expressed with the traditional zodiacal light brightness 
values given in $S_{10}(V)$, i.e. the equivalent number of 10\textsuperscript{th}
visual magnitude solar type stars per square degree. 
This unit was historically useful since the zodiacal light presents a solar spectrum in 
the visible wavelengths and so was valid over the visual domain. 
While SI units, i.e., W~m$^{-2}$~sr$^{-1}~\mu$m$^{-1}$ are easier to understand, 
they nevertheless need to be associated to a precise wavelength 
\citep[1 $S_{10}(V)$ is equal to $1.261\times 10^{-8}$~W~m$^{-2}$~sr$^{-1}~\mu$m$^{-1}$ 
at 550~nm, see][]{levasseur-regourd2001b}.

The brightness follows the power-law variation given in Table~\ref{zod:tab1} and 
is consistent with the local dust density distribution under a homogeneous 
zodiacal cloud hypothesis as described in \citet{leinert1977}.
However, taking into account the variation in albedo with respect to the heliocentric
distance gives a local dust particle density value following a power-law with 
an exponent of $-0.93 \pm 0.07$ (significantly 
different from the -1.3 value obtained under the homogeneity assumption)
\citep{levasseur-regourd1991a}. The Poynting-Robertson effect predicts 
a 1/R dependence for the density of dust particles in circular orbits located 
within the region of production of dust particles (ranging from 0.1~au to about 10-20~au), 
which is consistent with 
the power-law coefficient thus determined \citep{hanner1980}. 

The degree of linear polarization, $P$, variation  includes data obtained from a compilation 
by \citet{fechtig1981}, data inverted by \citet{levasseur-regourd1996}, data from the 
\citet{dumont1975} survey and data observed close to the Sun inverted by 
 \citet{mann1998}. 
Numerical simulations from \citet{lasue2007} and laboratory 
measurements from \citet{hadamcik2018a} are also superposed to the observations. 
$P$ decreases with decreasing solar distance. 
The sharp decrease noticed below 0.3~au suggests 
that drastic changes to the dust particles properties take place closer to the Sun, 
in agreement with trends obtained in the F-corona \citep{mann1996}. 

The dust properties variation with the solar distance could be related to temporal 
evolution of the dust particles, with possible changes in their composition and/or 
their size distribution, as they spiral towards the Sun under the 
Poynting-Robertson effect and get progressively warmer. 
This will be discussed in Section~\ref{section:origin}.

\textbf{ Figure 7: Local brightness and polarimetric data}

\begin{figure}[t]
\begin{center}
 \includegraphics[width=0.8\textwidth]{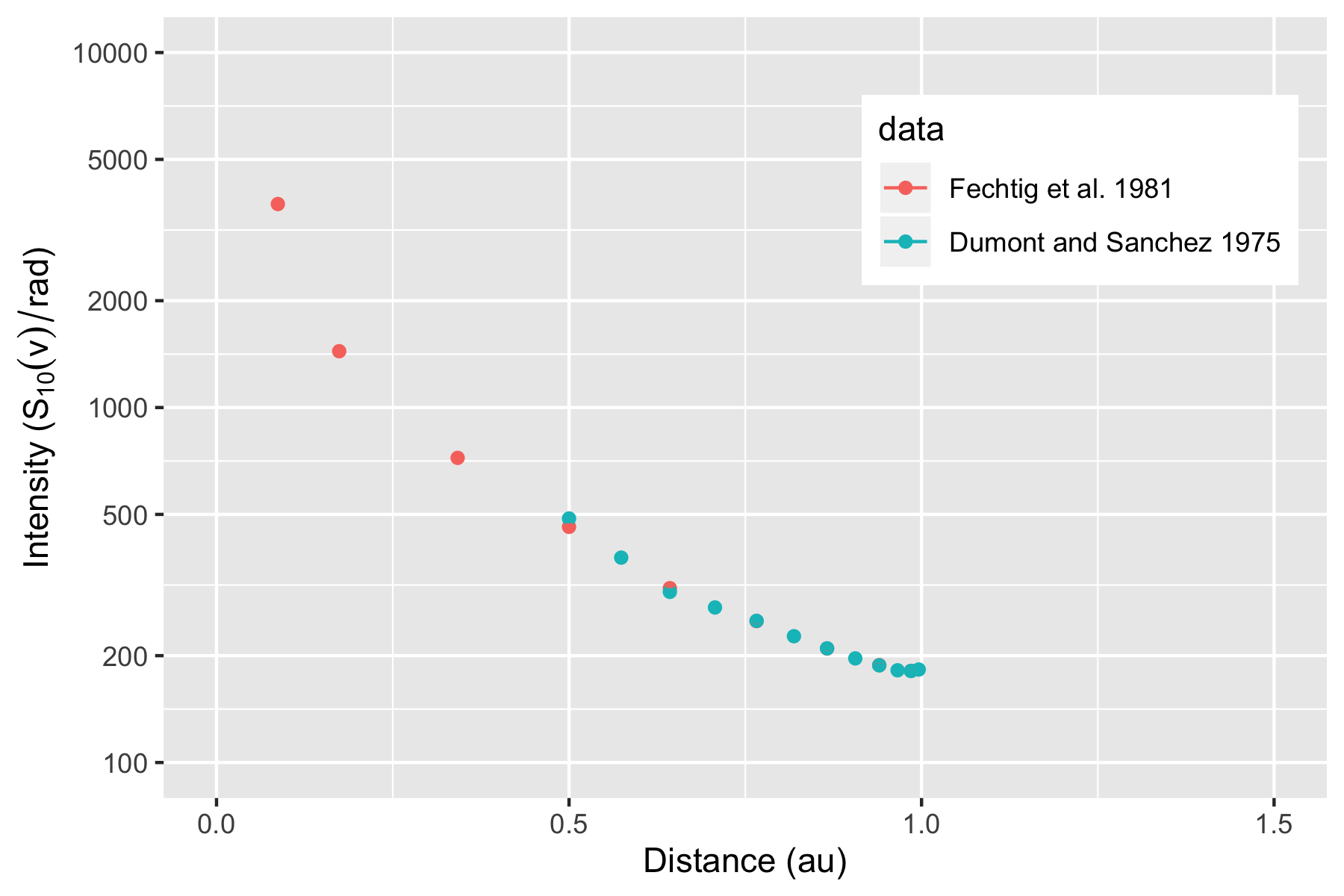} 
 \includegraphics[width=0.8\textwidth]{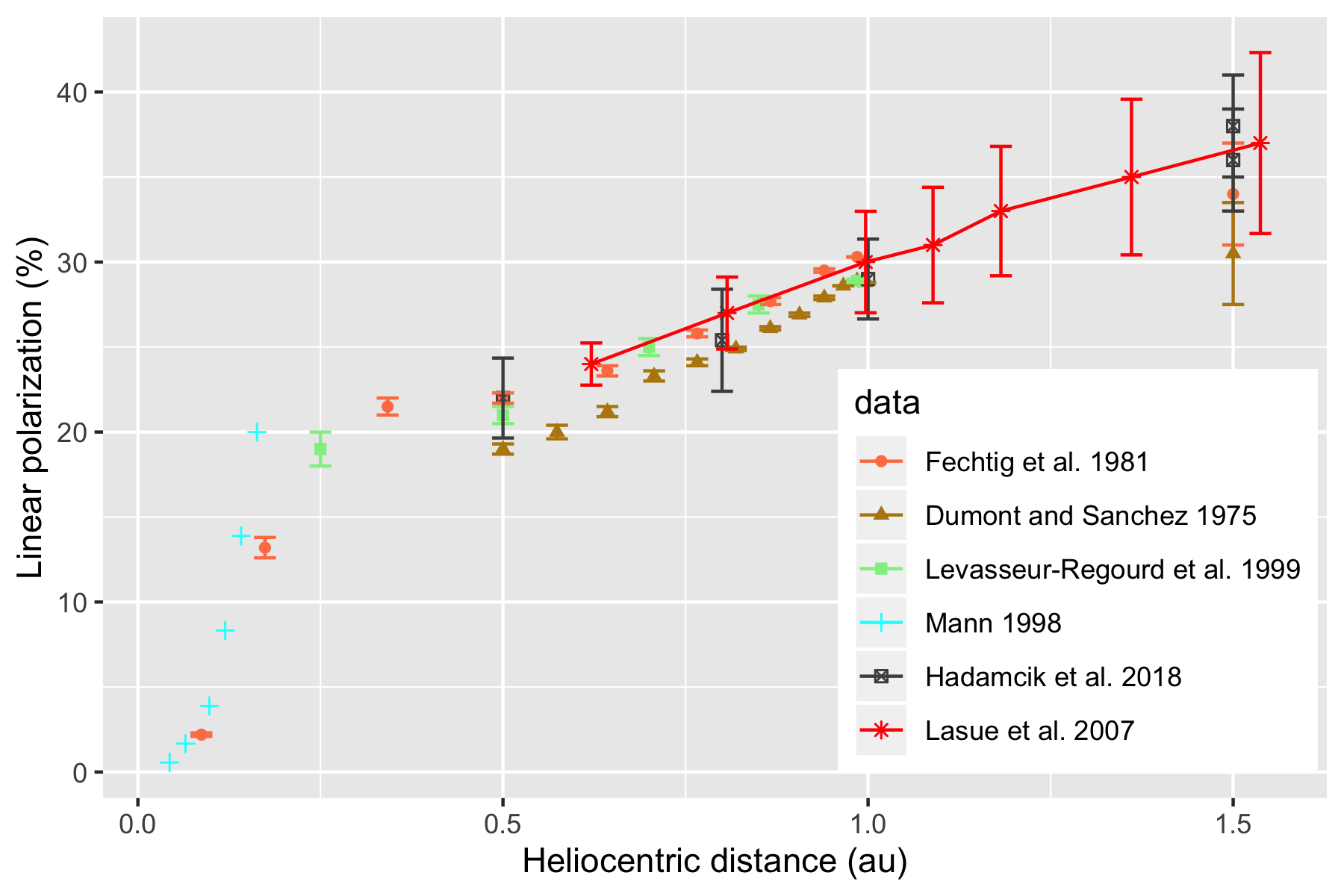} 
 \caption{Variation of the local near-ecliptic brightness (top) and  
 linear degree of polarization, $P$,  (in percent, bottom), at  $90^{\circ}$ 
 phase angle as a function of the heliocentric distance (au),
 adapted from  \citet{dumont1975, fechtig1981,levasseur-regourd1996, mann1998, 
 hadamcik2018a, lasue2007} and references within.}
  \label{zod:fig7}
\end{center}
\end{figure}

\subsection{Phase angle variations}

The local brightness and polarization phase curves of the zodiacal light at the martian node
($\approx1.5$~au) are plotted in Fig.~\ref{zod:fig7}. 
One can notice that around the backscattering domain ($\alpha < 20^{\circ}$), the intensity decreases
by $0.015\pm 0.05$ magnitude per degree. This demonstrates the existence of an opposition 
effect for the interplanetary dust population \citep{muinonen1991}.

The overall shape of the polarization phase curve is smooth, with a slightly 
negative polarization in the backscattering region, an inversion angle around $(15 \pm 5)^{\circ}$ 
and a slope at inversion of $(0.2 \pm 0.1)$ percent per degree. This shape is typical 
of light interaction with irregular dust particles possibly in the shape of 
fluffy aggregates made of submicronic monomers \citep{levasseur-regourd1997, lumme1997}.
The behavior of both intensity and polarization phase curves is comparable to 
that of cometary dust or C-type asteroids. 

The slope at inversion of the polarimetric phase curve can be related to the geometric albedo 
by the `Umov empirical law' \citep{umow1905}. From the inverted data, a geometric albedo value 
of $0.15 \pm 0.08$ can be calculated at 1~au \citep{levasseur-regourd1998}, which is consistent
within the error bars with the value given in Table~\ref{zod:tab1} and 
also consistent with the more recent analysis done by \citet{ishiguro2013}.

Finally, inversion of the local properties of the dust particles out of the symmetry plane 
of the zodiacal cloud in the plane perpendicular to the ecliptic and tangential to Earth’s orbit, 
have also been performed and normalized to the data inverted 
at 1~au and for $\alpha=90^{\circ}$. The brightness decreases with the elevation, indicating 
a flattened geometry for the interplanetary dust cloud. The polarization also decreases 
with elevation by about 10\% at 1~au while the albedo increases simultaneously 
from 0.8 to 0.11 \citep{renard1995}. 
These variations can be explained by the presence of two contributing dust populations, 
one coming from the asteroid belt and short period comets activity close to the ecliptic, 
and the other one, more isotropic, and  originating from long period comets and with 
different properties.


\textbf{ Figure 8: Local brightness and polarimetric data}

\begin{figure}[t]
\begin{center}
\includegraphics[width=0.8\textwidth]{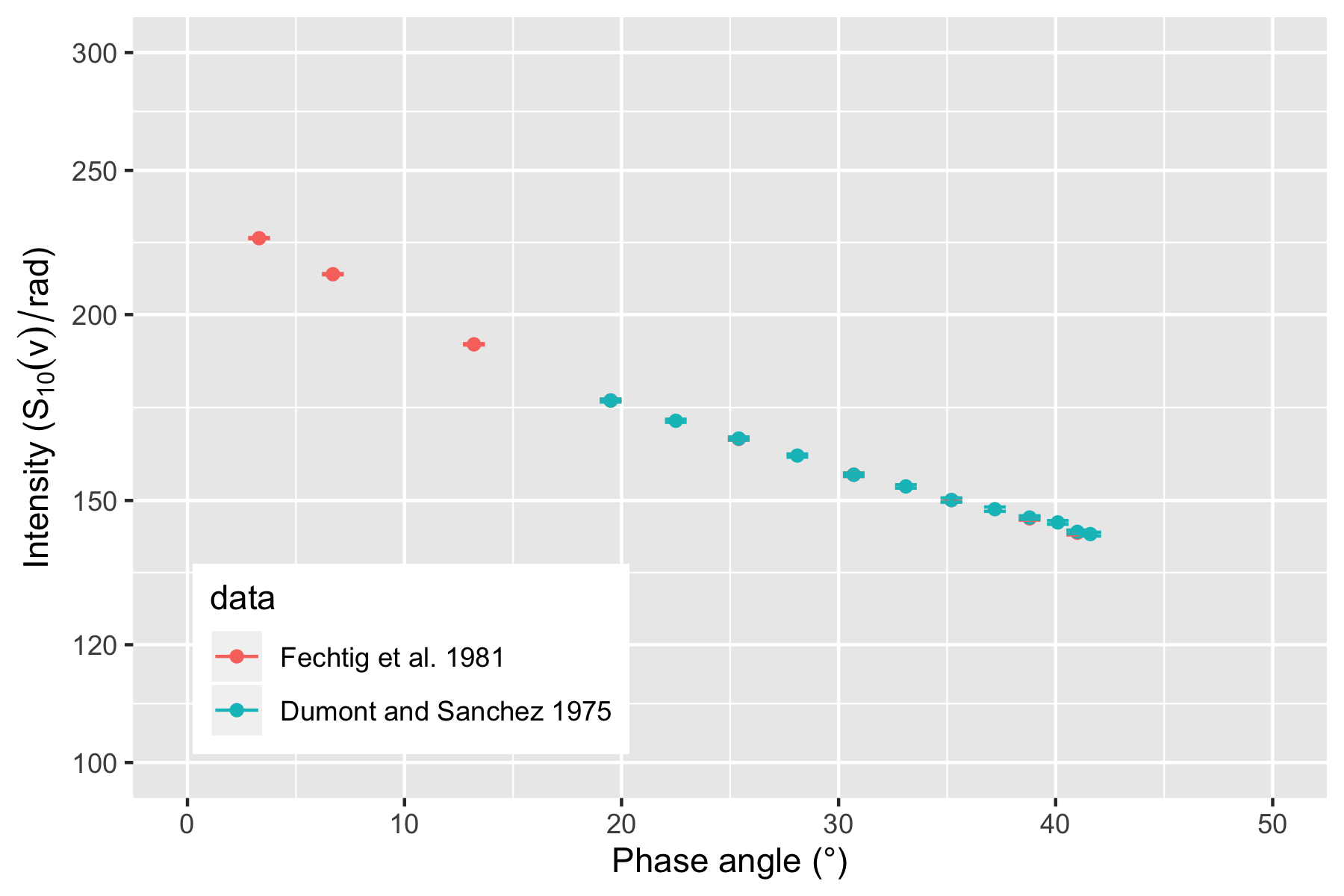} 
\includegraphics[width=0.8\textwidth]{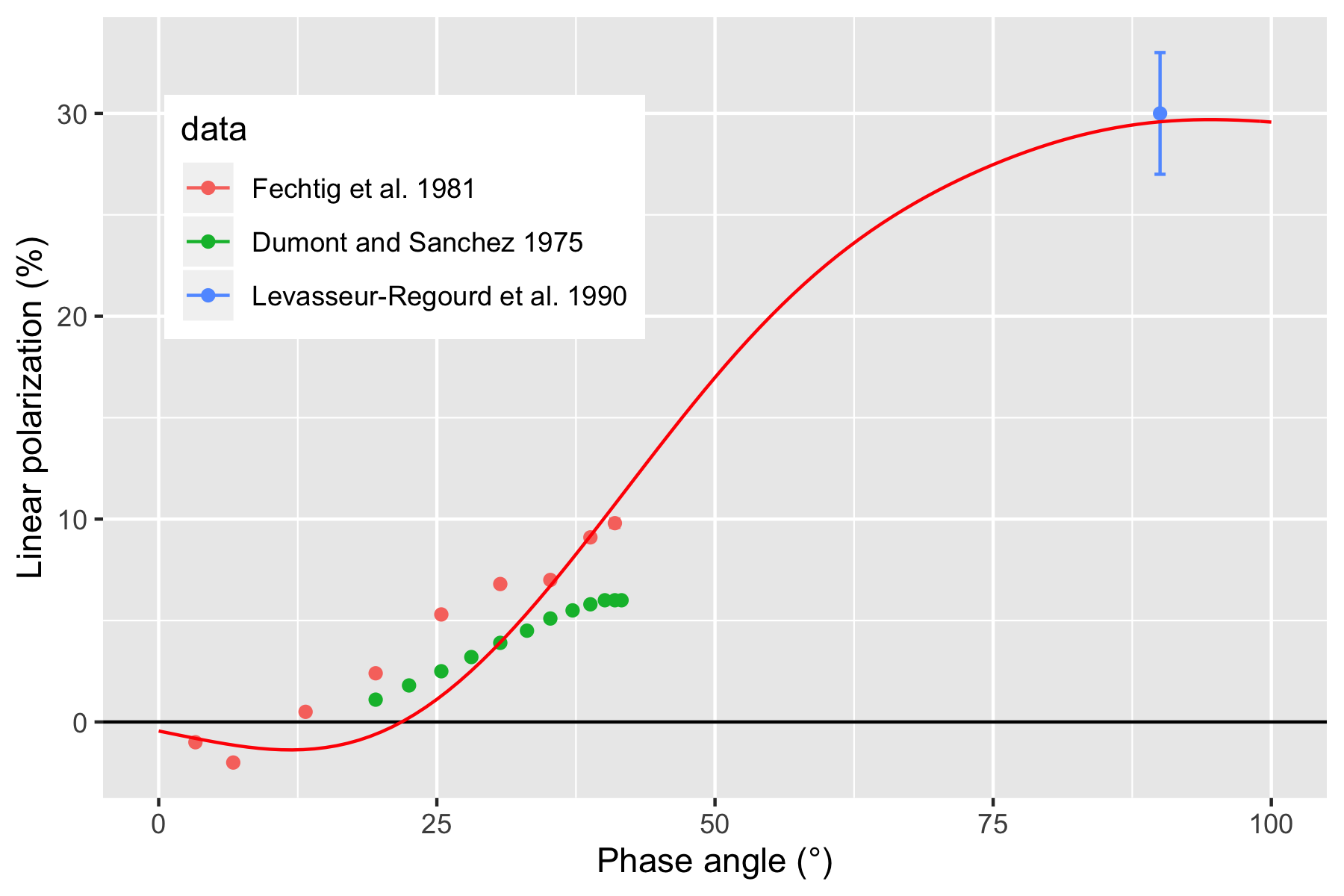} 
 \caption{Phase curves of the local near-ecliptic brightness (top) and  
 linear degree of polarization, $P$,  (in percent bottom), at 1.5~au,
 adapted from  \citet{dumont1975, fechtig1981,levasseur-regourd1990}  and references within.}
  \label{zod:fig8}
\end{center}
\end{figure}

\section{On the origins of interplanetary dust particles}
\label{section:origin}

\subsection{Numerical and laboratory simulations}

The intensity and polarization phase curves can be compared to numerical  
and experimental simulations in order to retrieve the likely physical properties of the scatterers. 

A classical assumption, based on astronomical observations, 
is that interplanetary, cometary and interstellar dust particles consist
of silicates, carbonaceous compounds and possibly ices 
\citep[see e.g.][]{greenberg1990a, ehrenfreund2004a}. 
The optical indices of such materials would present both the possibility 
for high absorption (carbonaceous compounds and dirty ices) and for low absorption
of light (pure ice and some silicates). 
Light scattering by irregular solid particles, especially micrometer-sized fluffy aggregates of 
submicrometer grains made of such materials are typically consistent with the intensity and 
polarization phase curves observed from interplanetary dust 
\citep{giese1978, haudebourg1999, nakamura1999}.  
Furthermore, a red polarimetric color variation of the scattered light by irregular particles
is linked to a lower porosity of the dust particles. It is therefore expected that compact
dust particles, possibly in the shape of elongated spheroids to present less symetries 
than spheres, could be significant contributors
to the interplanetary dust distribution \citep{lasue2007, kolokolova2010}. 

As shown in Fig.~\ref{zod:fig7}, the variation in local scattered intensity and 
polarization gives information on the heliocentric changes in dust density and  
properties. To the first order, the scattered intensity can be explained by the dust 
density variation consistent with the Poynting-Robertson effect. 

From numerical simulations  in \citet{lasue2007}, 
that have been confirmed by laboratory simulations \citep{hadamcik2018a}, 
the decrease in polarization sunward around 1~au can be explained to first order
by a change in the optical index of the dust particles population scattering 
the light. In that hypothesis, the proportion of absorbing materials, such as 
carbonaceous compounds,  to weakly 
absorbing ones would decrease sensibly towards the Sun 
(from 50-100\% at 1.5~au to about 0\% at 0.6~au).  A plot of the numerical simulations
and experimental measurements data points are shown on Fig.~\ref{zod:fig7}b.
This explanation would be consistent with an increase in the dust population albedo sunward, 
which is expressed by the variation equation shown for the albedo in Tab.~\ref{zod:tab1}.
Other changes in the dust population properties may also occur. 

Still in Fig.~\ref{zod:fig7}, closer to the Sun, below 0.3~au, one notices a sharp 
decrease in polarization as well as a change of slope in the scattered intensity. 
This change in behavior indicates that some properties of the dust cloud 
(size distribution, morphology, composition, etc.) is abruptly changing in that region. 
Many materials may sublimate in that  region and are certainly contributors
to changes in either size distribution, shape or optical indices of the dust 
particles \citep{mann2004}.


As interplanetary dust particles spiral inwards and vanish in the Sun, with a typical 
lifetime of a couple of million years at 1~au \citep{burns1979}, the interplanetary dust 
cloud requires constant replenishment. Comet nuclei activity as well as asteroidal 
collisions in the Main Belt can provide substantial amounts of dust particles and 
have long been considered to be major contributors to the inner zodiacal cloud
\citep{whipple1950a, dermott1984}. Other minor contributions come from the giant 
planets environment, the Kuiper Belt, and from interstellar dust particles 
\citep[see e.g.][]{grun1993}.
Observations of the visual zodiacal intensity within 1~au by the Clementine space probe
fitted with a model of dust distribution including asteroids, 
Jupiter Family Comets (JFC), Halley-type comets and Oort cloud comets indicates that 
at least 89\% of the interplanetary dust located in a sphere within 1~au from the Sun 
originate from comets \citep{hahn2002}.
Numerical simulations of the IRAS observations in the thermal domain 
of the zodiacal cloud based on the dynamical properties of the dust ejected 
from comets and from asteroids indicates that 85-95\% of the mid-infrared emission 
visible from 1~au is produced by  JFCs, while 
asteroidal and long-period comets are both respectively smaller than 10\%
\citep{nesvorny2010b, nesvorny2011}. 
Similar conclusions are reached with the dynamical study of the 
evolution of aggregate dust particles injected in the inner solar system by
active comets and with a size frequency distribution like 
the one of 67P/Churyumov-Gerasimenko, as they can reproduce the interplanetary 
dust properties detected around the Earth's orbit \citep{yang2018}.
Consistent results were also obtained from the study of the Doppler shift
from the Fraunhofer lines of the scattered solar light \citep{ipatov2008}.
An improved model of the IRAS and COBE 
thermal emission observations by \citet{rowan-robinson2013c} determined similar numbers
with, in addition, less than 1\% interstellar dust contribution at 1~au. 
Finally, lidar observations of Na and Fe fluxes in the Earth's atmosphere coupled with 
the cosmic spherule accretion rate measured at the South Pole also indicates that 
JFCs contribute $87\pm17\%$ of the total extraterrestrial matter mass input 
on the Earth \citep{carrillo-sanchez2016a}.

In summary, based on those diverse lines of evidence, and others described in the previous
sections, a consensus appears to emerge that the majority
of the dust particles located within 1~au should be contributed from the activity
of comets, and in particular the JFCs. 

\subsection{Properties of interplanetary dust and their link with comets}

As we have seen above, the dust particles at the origin of the zodiacal light 
are expected to present specific characteristics based on the properties 
of the scattered and emitted zodiacal light in the visible and the thermal 
wavelengths. Several lines of evidence based on different arguments make 
a compelling case to link the properties of the interplanetary dust particles 
to those of cometary dust particles. 

First, the dust density distribution models and related dynamics 
indicate that the majority of dust particles within 1~au from the Sun probably 
originate in major part from the activity of short period comets. 
This observation is also consistent with some of the properties detected for the
meteoroids population impacting the Earth. For example, almost all meteor showers 
have been linked with cometary parent bodies \citep{jenniskens2006}. 
The properties of interplanetary dust particles collected in the stratosphere of the Earth 
by airplanes are also consistent with primitive material and represent a family of samples that 
are chemically unlike the extraterrestrial matter represented by the meteorites 
(originating from asteroids apart from possibly one or two exceptions) 
\citep{flynn2016, koschny2019}. 
A population of micrometeorites named as UCAMM (Ultra-Carbonaceous Antarctic 
MicroMeteorites) for their very high carbon content (from 48\% to 85\%)
has also been studied and present a D/H ratio more than one order of magnitude larger 
than the terrestrial values strongly suggesting a cometary origin \citep{duprat2010}.

Second, the properties of the zodiacal light are consistent with a mixture 
of irregular fluffy particles and compact particles formed of silicates and 
carbonaceous materials. These properties are consistent with the properties of interplanetary 
dust collected in the stratosphere \citep{flynn2016}. 
They are also coherent with the properties 
of cometary dust particles as observed from the ground, analyzed in situ by 
sample return from Stardust and more recently from the in situ analysis performed 
by the Rosetta space mission \citep[see e.g.][]{levasseur-regourd2018b}. 
Most notably, the presence of both fluffy aggregates of grains and compact particles 
have been evidenced by both Stardust and Rosetta space missions 
\citep{horz2006b, guttler2019}. Similarly, both the Stardust samples and the Rosetta 
measurements have shown a distinct population of carbonaceous material 
forming a major fraction of the ejected dust with characteristics different from 
meteorites and possibly atmospheric IDPs \citep{sandford2006, fray2016b}.  
Finally, the scattered light phase curves present a similar behavior 
between ejected cometary dust and interplanetary dust particles analogues 
\citep{levasseur-regourd2019}.

Overall the study of the zodiacal light can give us access to properties 
of the cometary particles that may not be accessible by other means, since samples 
that arrive on Earth may be biased, and cometary space missions are limited in 
numbers. Continued monitoring of the zodiacal light over time will provide 
a better understanding of the origin and evolution of the interplanetary dust 
environment. 
Going one step further, the study of the similarities between the light scattering 
behavior of interplanetary dust particles, asteroids, 
cometary dust clouds and protoplanetary disks will help us understand planetary formation 
processes and the origin of the solid particles in these different environments 
\citep{levasseur-regourd2020}.

\section{Conclusions and future observations}

In this work, we attempted to give an overview of the properties of the zodiacal light 
and its link with the physical properties of the interplanetary dust particles that evolve 
in the Solar System and their sources. 
The zodiacal light observations provide us with an overall picture of the 
interplanetary dust environment: its mass flux distribution; chemical, material 
and structural properties; and the importance of different cometary and asteroidal sources in 
populating the inner Solar System with dust. 
This information has implications for the determination  
of the meteoroid flux deposited over the planetary surfaces. 

Intensity and polarization values of the zodiacal light are also of major importance 
for outer Solar System observations, since they provide an estimate of the foreground noise 
induced by the zodiacal light, together with an optimization of the epochs of observations 
of faint extended astronomical objects \citep{leinert1998, levasseur-regourd2019a}

We can expect the next generations of sensitive whole sky surveys 
to provide many additional constraints on the zodiacal light  with the possible 
detection of additional local heterogeneities and global time variability. 
Such surveys will include ground-based telescopes such as the 
Large Synoptic Survey Telescope (LSST), an 
optical 8.4m telescope designed to survey the visible sky every week starting in 2020 
\citep{abell2009}, but also space missions such as the Messier mission, 
a space telescope that will scan the night sky in the 200-1000 nm range, 
reaching surface brightness levels of 34 and 37 mag arcsec$^{-2}$ in 
the UV and optical domains \citep{valls-gabaud2016, muslimov2017}.
These results will need to be compared with the results obtained by 
the space missions studying the small bodies of the Solar System, 
such as  Hayabusa2 and OSIRIS-REx missions for asteroids 
and the newly selected ESA Comet Interceptor mission to flyby 
a new pristine active comet \citep{gater2019}. Further data out of the 
Earth's orbit could be obtained closer to the Sun from the recently launched 
Parker Solar Probe, and by other probes tangentially to the orbits of planets
or small bodies they may be orbiting. 
Also, future observations of the zodiacal light may be sensitive to 
circular polarization, which could constrain the origin of the particles and specific dynamic 
properties in the solar magnetic field.

Some important unsolved questions related to the zodiacal light that need 
to be studied in the future include:  
\begin{itemize}
\item{the exact origin and physical properties of the dust particles contributing to the zodiacal cloud, }
\item{the possible time evolution of the zodiacal cloud over decades, }
\item{ the circular polarization that can probe the interplanetary magnetic field structures, }
\item{the dust evolution in  the vicinity of the Sun.}
\item{how the properties of the zodiacal cloud could help us constrain the properties
of exo-zodiacal dust clouds, which may behave quite differently depending on 
the presence and the properties of orbiting exoplanets.}
\end{itemize}

Answers will progressively be found as future observations continue to focus 
on higher resolution photopolarimetry of the zodiacal light at all wavelengths 
and possibly observed out of the ecliptic.
 Improvements on the inversion methods based on such measurements
will  also likely further improve zodiacal light studies.

\section{Acknowledgements}

This work was supported by the Programme National de Plan\'etologie
(PNP) of CNRS/INSU, co-funded by CNES.
We acknowledge the organizers of the Meteoroids 2019 conference for inviting 
the authors to present this work. 
We thank two anonymous referees for providing interesting discussions and improvements 
on the manuscript.

%
%


\bibliography{2019_PSS_zod_review}

\end{document}